\documentstyle[aaspp4,epsfig]{article}
\def\simlt{\lower.5ex\hbox{$\; \buildrel < \over \sim \;$}}
\def\simgt{\lower.5ex\hbox{$\; \buildrel > \over \sim \;$}}
\def\simpt{\lower.5ex\hbox{$\; \buildrel \propto \over \sim \;$}}
\def\mag{\mbox{ mag}}
\def\kms{\mbox{ km s$^{-1}$}}
\def\mpc{\mbox{ Mpc}}
\def\kpc{\mbox{ kpc}}
\def\pc{\mbox{ pc}}

\def\msun{\mbox{ M}_\odot}

\def\civ{C\,{\scriptsize IV}~}

\begin{document}
%% \special{!userdict begin /bop-hook{gsave 200 30 translate 65 rotate 
%% /Times-Roman findfont 220 scalefont setfont 0 0 moveto 0.93 setgray 
%% (DRAFT) show grestore}def end}

\title{Confronting $\Lambda$CDM with Gravitational Lensing Constraints
  on Small Scale Structure}

\author{R. Benton Metcalf\footnote{Hubble Fellow}}
\affil{\it Department of Astronomy and Astrophysics, University of California,
Santa Cruz, CA 95064 USA}

\abstract{
This paper primarily addresses the question of whether recent lensing
observations probing the small scale structure in the universe are
consistent with the $\Lambda$CDM model.
A conservative approach is taken where only the most difficult to explain
cases of image flux anomalies in strong lenses are considered.
Numerical simulations are performed to 
compare predictions for the $\Lambda$CDM small scale mass function
with observed flux ratios.
It is found by simulating several represent cases that all the cusp
caustic lens anomalies and the 
disagreements between monochromatic flux ratios and simple lens models might be 
explained without any substructure in the primary lenses' dark matter
halos.  Intergalactic $\Lambda$CDM halos are enough to naturally explain these
cases.  However, thus far, spectroscopic gravitational lensing observations
require more small mass halos ($\sim 10^6\msun$) than is expected in
the $\Lambda$CDM model. 
}

\section{Introduction}
\label{sec:introduction}
The Cold Dark Matter (CDM) model predicts a large quantity of small mass dark
matter halos ($\simlt 10^7 \msun$) that must have little or no stars in
them to agree with the number counts of dwarf galaxies.  Quasars
(QSOs) that are being gravitationally lensed into multiple images have
recently been used to put limits on the surface density and mass of such invisible
subclumps 
\markcite{1998MNRAS.295..587M,2001ApJ...563....9M,2002ApJ...565...17C,2002ApJ...580..696M,2002ApJ...567L...5M,Dalal2002,astro-ph/0112038,2003ApJ...584..664K,cirpass2237}({Mao} \& {Schneider} 1998; {Metcalf} \& {Madau} 2001; {Chiba} 2002; {Metcalf} 2002; {Metcalf} \& {Zhao} 2002; {Dalal} \& {Kochanek} 2002; {Brada{\v c}} {et~al.} 2002; {Keeton} 2003; {Metcalf} {et~al.} 2004).
Small mass clumps near the images affect the observed magnifications
ratios.  The question arises as to whether these observations are
compatible with the current $\Lambda$CDM model.

This question is significantly complicated by the fact that all
lenses were not created equal.  Some lenses provide much stronger and
more certain constraints on the small scale structure than others.  In
this paper, I try to take a conservative 
approach and consider only the lenses that provide clean, relatively
unambiguous constraints.  I also refrain from doing a formal
likelihood analysis to constrain structure formation parameters 
because I think this would be premature considering the uncertainties
in the relevant $\Lambda$CDM predictions and the small amount of data
at this time.

In this paper, the single large lens that is causing the QSO to have
multiple images is referred to alternately as the primary lens, the host
lens or the host halo.  The additional small scale halos are referred
to as subhalos or substructures even if they are not physically
inside the host halo, but in intergalactic space.
For the purposes of this paper the standard $\Lambda$CDM cosmological
model will have the cosmological parameters $\Omega_m=0.3$, 
$\Omega_\Lambda=0.7$, $\sigma_8=0.9$, $H_o=70 \kms\mpc^{-1}$ and a scale
free initial power spectrum.

In section~\ref{sec:expect-lambd}, the predictions of the $\Lambda$CDM
model are discussed.  Relevant background information about strong
gravitational lensing and the techniques used to probe substructure
are reviewed in section~\ref{sec:some-lens-backgr}.  A brief summary
of relevant observations is in
section~\ref{sec:summary-observations}.
Section~\ref{sec:simulations} provides a description of the lensing
simulations.  The results of the simulations are compared with the
observations in section~\ref{sec:results} and
in section~\ref{sec:discussion} the importance of these results are discussed.

\section{Expectations for $\Lambda$CDM}
\label{sec:expect-lambd}
Cosmological Nbody simulations predict that $\sim 10-15\%$ of the mass
within the virial radius of a $10^{12}\msun$ halo is in substructures
with $m \simgt 10^7 \msun$
\markcite{1999ApJ...524L..19M,1999ApJ...522...82K}({Moore} {et~al.} 1999; {Klypin} {et~al.} 1999).  Cosmological simulations are
limited to particle masses of $\simgt 10^6\msun$ so smaller substructures
cannot be probed directly.  
For the strong lensing studies considered here, we are interested in the
mass fraction in substructure at a projected radius of $\sim 10\kpc$
which may be
substantially less than the value for the halo as a whole because of
tidal stripping, tidal heating, and dynamical friction.  Limited
resolution can make overmerging a problem at these radii.  The lensing
observations are also sensitive to substructure masses well below the
resolution of the simulations.  In addition, baryons may play a
significant role in determining the structure of the halo at these small
radii and no simulation has yet fully incorporated them at high enough
resolutions.  As a result of these complications, the predictions of
$\Lambda$CDM as they pertain to substructure in strong lenses are not
certain.  They must be extrapolated from the simulations of
insufficient resolution.

\markcite{2004ApJ...604L...5M}{Mao} {et~al.} (2004) have done Nbody simulations in an
effort to determine the level of substructure.  They find that 
$\simlt 0.5\%$ of the surface density at appropriate projected radii
is in structures with $m \simgt 10^8\msun$.  It is uncertain how
accurate this estimate is since no thorough convergence tests have
been done in this regime.
In addition, below this mass dynamical friction becomes considerably less
effective \markcite{2001ApJ...559..716T,2004MNRAS.348..811T}(see {Taylor} \& {Babul} 2001, 2004).  Dynamical friction erodes the orbits of large satellite
halos, causing them to be destroyed as they sink to the center of
their host halo.
\markcite{2004MNRAS.348..333D}{De Lucia} {et~al.} (2004) have also studied halo substructures for
masses $\simgt 10^9\msun$ and find that the mass function is
independent of the host halo mass.

\markcite{astro-ph/0304292}Zentner \& Bullock (2003) have developed a method for extrapolating the
results of Nbody simulations to smaller masses and radii.  Using their
figure~19 it can be estimated that the fraction of the surface density in
satellites of mass $10^5\msun < m < M_{\rm sat}$ is
\begin{equation}\label{sub_frac}
f_{\rm 10kpc} \simeq 0.01 \left(\frac{M_{\rm
    sat}}{10^{9}\msun}\right)^{0.5}
\end{equation}
(for $10^6\msun \simlt M_{\rm sat}<10^9\msun$) 
at a projected radius of 10~kpc which is appropriate for the strong
lenses considered here.  Almost all of these subhalos are more than
$30\kpc$ -- or several times the typical Einstein radius -- from the center of the host halo in 3 dimensions.  Analytic models have also been constructed
by \markcite{2004MNRAS.348..811T}{Taylor} \& {Babul} (2004) who claim that Nbody simulations may be
suffering from overmerging at small halo-centric radii 
\markcite{astro-ph/0312086}(see also Taylor, Silk, \& Babul 2004).  They argue that because of this
the above might underestimate the substructure mass function by a factor of several.
However, they do not provide a prediction that can be easily compared to
the lensing.  For definiteness, equation~(\ref{sub_frac})
will be considered the $\Lambda$CDM predication for substructure inside
the primary lens in this paper.   In this sense the Nbody results, and
extrapolations of them, are taken at face value although it is still 
possible that these simulations do not accurately reproduce the
$\Lambda$CDM model in this regime.  For example, the role of baryons
is not taken into account.

In addition to the substructure inside the host lens there are also
independent halos in intergalactic space that happen to be well aligned
with the source, lens and observer.   The number of these halos can
be calculated straightforwardly using the Press-Schechter \markcite{1974ApJ...187..425P}({Press} \& {Schechter} 1974)
method and the Sheth-Tormen \markcite{2002MNRAS.329...61S}({Sheth} \& {Tormen} 2002) modification
to it.  A typical line of sight to $z=2$ passes within 1/3
of the virial of 150 halos of mass $10^5\msun < m <10^9\msun$.  Since
the deflections from these halos will add, they can make a 
contribution to the lensing that is significantly larger than one
halo could do by itself.   We will see that they
have an important effect on the magnification of any small source at
high redshift.

Besides the mass function of halos one must also consider how the
concentration of the halos depends on mass.  The Nbody simulations
are generally not of high enough resolution to determine the
concentration of halos with masses below $\sim 10^9\msun$ that are inside
the halos of large galaxies.
Some progress can be made in this regard by dropping ``live''
artificially constructed satellites into a static model for the host
halo extracted from a cosmological simulation \markcite{2003ApJ...584..541H}(as
  in {Hayashi} {et~al.} 2003).  The subhalos are taken to have 
Navarro, Frenk \& White (NFW) profiles \markcite{1997ApJ...490..493N}({Navarro}, {Frenk}, \&  {White} 1997)
\begin{equation}
\rho(r)=\frac{r_s\rho_o}{r(1+r/r_s)^2}
\end{equation}
The simulation results indicate that
substructures are effectively tidally truncated at some radius with the
interior remaining relatively unmodified until the stripping radius
becomes on the order of the scale length, $r_s$.  This is the simple picture
that will be used for the simulations in this paper.  By extrapolation
of Nbody simulations \markcite{astro-ph/0304292}Zentner \& Bullock (2003) find that
the concentration of small halos goes as 
\begin{equation}\label{concentration}
c \equiv \frac{r_{\rm vir}}{r_s}\simeq c_o \left(\frac{m_{\rm
    vir}}{10^{12}\msun}\right)^{-\beta} 
\end{equation}
with $c_o\simeq 12$ and $\beta\simeq 0.10-0.15$.  In this paper
$\beta=0.13$ is adopted.  $m_{\rm vir}$ is the virial mass of the subhalo before it
is tidally stripped.

\section{Some Lensing Background}
\label{sec:some-lens-backgr}
\begin{figure}[t]
\centering\epsfig{figure=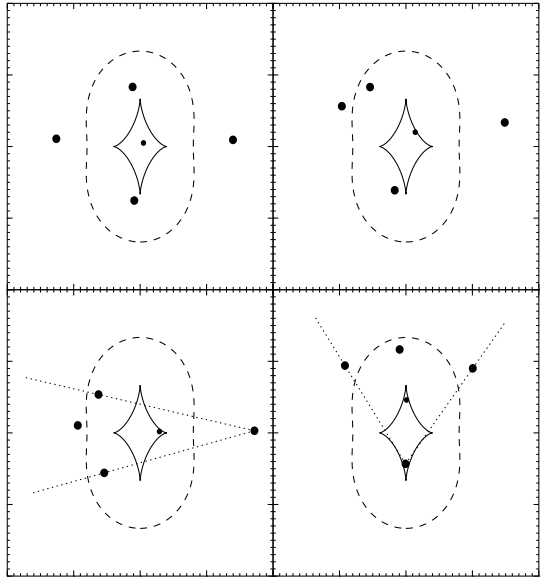,height=3.0in}
\caption[the]{\footnotesize Four basic lens configurations.  In each
  case the caustic is shown as a solid curve and the critical curve is
  shown as a dashed curve (only one of each for each configuration is shown).  The four images
  that are usually observable are shown as large dots and the source
  position is marked by a small dot.  On the top left is the {\it Einstein
  cross} configuration where all the images are well separated and the
  source is near the center of the lens which is at the center of each
  plot. On the top right is the {\it fold caustic} configuration where
  two of the images are close together and the source is near the
  caustic, but not near a cusp. The lower left shows a {\it short axis
  cusp caustic} configuration and the lower right is a {\it long axis
  cusp caustic} configuration.  The {\it image opening angle} is the angle
  between the dotted lines shown in the cusp caustic cases.  Note that
  this opening angle is defined differently here than it is in some other papers
  where the center of the lens is taken as the vertex.  There are
  always two images within the critical curve where the magnification is
  negative and two outside of the curve where the magnification is
  positive.  The long and short axis cusp caustic cases differ in that the close
  triplet of images have either one (long axis) or two (short axis)
  negative images.  They also differ in how close the singlet image is
  to the center of the lens which can usually be determined observationally.}
\label{fig:lens_diagram}
\end{figure}

Some background on strong gravitational lensing will be necessary to
understand the results that follow.  For a more complete description
see \markcite{SEF92}{Schneider}, {Ehlers}, \& {Falco} (1992), or any other review of strong lensing (see
\markcite{2003AJ....125.2769S}{Saha} \& {Williams} (2003) for
  a nice qualitative description).  A strong
lens can be defined as one where there are multiple images of a
single source.  For any lens -- that is less concentrated than a
perfect point mass -- there will be one image if the source is far enough
away from the center of the lens.  On the source plane of a potential
strong lens there are also regions where there are three images and,
when the lens is not perfectly axisymmetric, five images.  
One of these images is usually
near the very center of the lens and, if the density profile is very
cuspy there, this image is highly demagnified; in the
large majority of cases it is not observed 
\markcite{2004Natur.427..613W}(for an exception see {Winn}, {Rusin}, \&  {Kochanek} 2004).  This leaves two or four images.
Separating these regions on the source plane are the caustic curves.
If the source moves from outside a caustic to the inside of it two
images are created.  Generally for a smooth centrally concentrated
lenses there are two caustic curves -- termed the radial and
tangential caustics.

Figure~\ref{fig:lens_diagram} illustrates the basic configurations
for four image lenses.  In this figure the central (or tangential)
caustic is shown as a solid curve and the critical curve that is the
image of the caustic curve is shown as a dashed curve.  Images within the
dashed curve have negative magnification reflecting the fact that
these images are reversed in one dimension with respect to images
that are outside the curve (i.e. negative parity in one dimension).
The two types of cusp caustic 
configurations differ in that for the long axis case the triplet
of close images includes one of these negative images and in the
short axis case it includes two.  The sign of the magnifications is
not directly observable, but for configurations other than
Einstein crosses one can usually deduce them because the
parities alternate from image to image as one follows the critical
curve and the positive parity images are generally further from the
center of the lens.  For example, the two types cusp caustic
configurations can be distinguished by how close the singlet image is
to the center of the lens.  To measure the degree of ``cuspyness'' the
image opening angle is defined as shown in figure~\ref{fig:lens_diagram}. 

To investigate the presence of substructure in a strong lens one must
find a prediction that is not strongly dependent on the macroscopic
form of the lens which is not known in detail.  The magnification
ratios are influenced by substructure \markcite{2001ApJ...563....9M}({Metcalf} \& {Madau} 2001),
but their values are model dependent which limits their use somewhat
and makes their interpretation ambiguous.  There are a few
observables that are relatively unambiguous.  They are discussed below.

\subsection{The cusp caustic relation}
\label{sec:cusp-caust-relat}
It can be proven by expanding the
lensing map to third order in the angular separation from a cusp
in the caustic that the magnifications of the close triplet of
images should sum to zero \markcite{1992A&A...260....1S}({Schneider} \& {Weiss} 1992).  To make 
this prediction independent of the intrinsic luminosity of the 
QSO the images in the triplet are labeled A through C and the cusp
caustic parameter, $R_{\rm cusp}$, is defined as
\begin{equation}
R_{\rm cusp}\equiv \frac{\mu_A+\mu_B+\mu_C}{|\mu_A|+|\mu_B|+|\mu_C|}
\end{equation}
which should be zero if the expansion of the lens map about
the cusp is valid.  Small scale structure on approximately the scale
of the image separations will cause $R_{\rm cusp}$ to differ from zero
fairly independently of the form of the rest of the lens.
By adding radial modes to analytic lens models \markcite{KGP2002}{Keeton}, {Gaudi}, \& {Petters} (2003) showed
explicitly that, for their family of lens models, $R_{\rm cusp}$ is
always small when the image opening angle is small and there are no
large fluctuations in the surface density on the scale of the image
separations.

Note that by the definition of $R_{\rm cusp}$ used here it can be both
negative and positive (some authors use the absolute value of $R_{\rm
  cusp}$).  Substructure is more likely to reduce the absolute
magnification than to increase it for negative magnification images 
\markcite{astro-ph/0109347,2001ApJ...563....9M,2002ApJ...580..685S}(Metcalf 2001; {Metcalf} \& {Madau} 2001; {Schechter} \& {Wambsganss} 2002).  
The positive parity images are biased in the other direction.  As a result, 
 the probability distribution of $R_{\rm cusp}$ will be skewed toward
 positive values.  We will see that this is a strong effect.  Also
 note that $|R_{\rm cusp}|<1$ by definition.

Cusp caustic systems also have the benefit that the time delays
between the images of the triplet are usually small, smaller than typical
time scales for the variations in the radio or infrared emission.
This makes the interpretation of the flux ratios as magnification
ratios more secure.

\subsection{Spectroscopic gravitational lensing}
\label{sec:spectr-grav-lens}

It was proposed by \markcite{MM02}{Moustakas} \& {Metcalf} (2003) that much of the lens model degeneracy
can be removed and the sensitivity to substructure properties
improved by utilizing the fact that the different emission regions of the
source QSO have different physical sizes.  If the lens is smooth on
the scales that bridge the sizes of the emission regions, the
magnification of those regions should be the same and thus the
magnification ratios should be the same.  The visible and
near-infrared (near-IR)
continuum emission regions are small, $\sim$ 100~AU
\markcite{2001ApJ...548L.127Y,2000MNRAS.318..762W,1990ApJ...358L..33W}({Yonehara} 2001; {Wyithe}, {Webster}, \&  {Turner} 2000; {Wambsganss}, {Schneider}, \&  {Paczynski} 1990),
and their magnification can be affected by microlensing by ordinary stars in the
lens galaxy.  The broad line emission region is $\sim 0.1\pc$ in size
\markcite{2000ApJ...533..631K,1999ApJ...526..579W}({Kaspi} {et~al.} 2000; {Wandel}, {Peterson}, \&  {Malkan} 1999) and is less affected
by microlensing in most cases.  The radio and 
mid-IR regions are $\sim 10\pc$
\markcite{1999MNRAS.306..161A,2002MNRAS.331.1041W}({Andreani}, {Franceschini}, \&  {Granato} 1999; {Wyithe}, {Agol}, \&  {Fluke} 2002) and their
magnification should be 
dominated by larger scales than stars.  The narrow line emission region
is even larger, $\sim 100\pc$ \markcite{2002ApJ...574L.105B}({Bennert} {et~al.} 2002).  The magnification ratios in these bands and lines
can be compared to constrain the mass, concentration and number
density of substructures \markcite{cirpass2237}({Metcalf} {et~al.} 2004).  A mismatch in the
magnification ratios can be expressed by the {\it differential
  magnification ratios} (DMR) which is formed by taking the flux ratio between
images for one emission region and then dividing by the flux ratio in
another emission region.  The DMRs will
all be 1 if there is no mismatch.  To further distill the
information, the {\it spread} is defined as the difference between the
largest DMR and the smallest DMR measured in magnitudes.  The spread
is independent of which image is used to normalize the ratios and
will be larger for larger the mismatch in the monochromatic flux ratios.

\subsection{bent radio jets}
\label{sec:bent-radio-jets}

Another idea for detecting substructure is to compare the images of a
radio jet in a strong lens \markcite{2002ApJ...580..696M}({Metcalf} 2002).  The Very Long
Baseline Interferometer (VLBI) is able to image these jets at
milliarcsecond resolution and in some cases can measure structures in
the radio jet.  Substructure can bend the jet in one image in a
different way than is seen in the other images.
In practice there can be some ambiguity in this kind of measurement because the
curvature of the jet in one image can be magnified in
another image by the host lens alone and, because of limited resolution, the
curvature of a jet is not often well measured.  However, the
relative directions of the image curvatures can be predicted in a
model independent way, i.e. the relative parities of the images can be predicted.  
A violation of this prediction would be an unambiguous signature of
substructure.  In general
this kind of observation is sensitive to substructures that are small
($m \sim 10^6\msun$) and strongly concentrated.

\section{Summary of Observations}
\label{sec:summary-observations}
At this time there are about 80 known gravitationally lensed QSOs with
multiple images.  A very useful resource for data on these lenses is
provided by the CfA-Arizona Space Telescope Lens Survey (CASTLES)\footnote{See
  http://cfa-www.harvard.edu/castles/ for a summary of current data.}
which is tasked with doing followup observations of all close 
QSO lenses in the visible and near-IR.
Of these prospective lenses some are two 
image lenses and some are cases where it has not yet been verified
that there is a single QSO being multiply imaged rather than multiple
QSOs.  Many of these 
lenses have been observed only at visible wavelengths or only at
radio wavelengths.  Only a small minority of them have sufficient
data to do a spectroscopic lensing study of them and/or are in a
configuration that makes the cusp caustic relation a significant
constraint.

There are several cases of particular interest here.  The data and
previous studies of these lenses are briefly summarized here.

\subsection{Q2237+0305}
\label{sec:q2237+0305}
This lens is probably the most well studied QSO lens.  It is in an
Einstein cross configuration with a lens redshift $z_{\rm lens}=0.04$ and
a source redshift $z_{\rm source}=1.69$.  Microlensing by stars has been
detected in this case through time variations in the magnification
ratios at visible wavelengths and used to study the structure of the
QSO \markcite{1989AJ.....98.1989I,2000ApJ...529...88W,2000MNRAS.318..762W,2002MNRAS.331.1041W}({Irwin} {et~al.} 1989; {Wo{\' z}niak} {et~al.} 2000; {Wyithe} {et~al.} 2000, 2002).

A spectroscopic lensing study of Q2237+0305 was done by
\markcite{cirpass2237}{Metcalf} {et~al.} (2004).  It was found that the broad line (H$\beta$),
mid-infrared, radio and narrow line ([OIII]) magnification ratios do
not agree (although the mid-infrared and radio ratios do agree which
is expected because of their similar size).  
The spread (see \S~\ref{sec:spectr-grav-lens})
%or largest discrepancy in the differential magnification ratios 
between the combined radio/mid-IR and the narrow lines is $0.77\pm 0.19\mag$.
It is shown that if substructures are responsible for this, they must have a
mass $10^5\msun\simlt m \simlt 10^8\msun$ and that their surface density
must be greater than 1\% of the total surface density of the lens for
typical assumptions about the radial profile of the lens and
substructures.  By comparison with equation~(\ref{sub_frac}) it can be
seen that this is in violation of the $\Lambda$CDM predictions.  Only
substructures within the primary lens were considered in
\markcite{cirpass2237}{Metcalf} {et~al.} (2004).  This study provides the strongest constraint on
the type, mass and concentration, of the substructures that could be
causing the magnification anomalies.

\subsection{B2045+265}
\label{sec:b2045+265}
This is the strongest case for a violation of the cusp caustic
relation.  The image opening angle is only $25.2^o$ making this an
extreme example.  The redshifts are  $z_{\rm lens}=0.87$ and $z_{\rm
  source}=1.28$ and it is a long axis cusp caustic configuration.  In the radio,
\markcite{1999AJ....117..658F}{Fassnacht} {et~al.} (1999) get $R_{\rm cusp}=0.516\pm0.018$ and 
\markcite{2003ApJ...595..712K}{Koopmans} {et~al.} (2003) get $R_{\rm cusp}=0.501\pm0.035$ after 14 measurements. 
\markcite{2003ApJ...595..712K}{Koopmans} {et~al.} (2003) have demonstrated that the fluxes
of the close triplet of images are varying independently at the 7\%
level (this is incorporated into the quoted error).  They attribute
this variation to scintillation within our galaxy.  However, it seems
unlikely that these variations are responsible for the large value of
$R_{\rm cusp}$ since the radio, near-IR and visible measurements all
agree (the CASTLES value is $R_{\rm cusp}=0.506\pm 0.013$).   

\subsection{B1422+231}
\label{sec:b1422+231}
This is the first case publicized as a violation of the cusp caustic relation
\markcite{1998MNRAS.295..587M}({Mao} \& {Schneider} 1998) and it has been further investigated in
this regard by a number of authors \markcite{Keeton2002,2002ApJ...567L...5M,astro-ph/0112038}(Keeton 2002; {Metcalf} \& {Zhao} 2002; {Brada{\v c}} {et~al.} 2002).  
The redshifts are  $z_{\rm lens}=0.34$ and $z_{\rm source}=3.62$.   
The configuration is a long axis cusp caustic with an image opening
angle of $61.0^o$ which makes it a less extreme case than B2045+265.
This lens has been observed in the radio by 
\markcite{2001MNRAS.326.1403P}{Patnaik} \& {Narasimha} (2001) and \markcite{2003ApJ...595..712K}{Koopmans} {et~al.} (2003) who
essentially agree on $R_{\rm cusp}=0.187\pm 0.006$ with no detectable
time variation.  The optical and near-IR measurements from CASTLES
are in agreement with this value.

\markcite{KGP2002}{Keeton} {et~al.} (2003) showed that the violation of the cusp caustic relation
in combination with the image opening angle is not in itself strong evidence for
substructure.  However, using explicit lens models, it
has been shown that it is difficult to 
construct a lens model for B1422+231 that fits the image positions,
resembles a realistic galaxy+halo and at the same time reproduces the
magnification ratios \markcite{2002ApJ...567L...5M,EW03}(for example {Metcalf} \& {Zhao} 2002; {Evans} \& {Witt} 2003).  

\subsection{B0712+472}
\label{sec:b0712+472}

This lens is a long axis cusp caustic case similar to B1422+231 in
that the image opening angle is $50.0^o$, but 
two of its images are significantly closer together than in B1422+231.
This indicates that the image is not located along the caustic cusp's
axis of symmetry (theoretically this does not affect the prediction
that $R_{\rm cusp}\simeq 0$).  The redshifts are  $z_{\rm lens}=0.41$
and $z_{\rm source}=1.34$.   The observed radio $R_{\rm cusp} = 0.26\pm 0.02$
\markcite{1998MNRAS.296..483J,2003ApJ...595..712K}({Jackson} {et~al.} 1998; {Koopmans} {et~al.} 2003).  The visible/near-IR
$R_{\rm cusp}$ is larger and a function of wavelength indicating that
differential extinction might be important at these wavelengths (see CASTLES).

\subsection{bent radio jets}
\label{sec:bent-jets-observed}

There are several cases where a distinct bend is visible in one or more of the
jet images.  One strong bend in lens MG0414+0534 is traceable to a visible dwarf companion
galaxy.  In addition B1152+199 has an unexplained mismatch in the image
curvatures that can be explained by substructure
\markcite{2002MNRAS.330..205R,2002ApJ...580..696M}({Rusin} {et~al.} 2002; {Metcalf} 2002).  
In this case the signal to noise in the measurement of the bend
is not large and the conclusion that the bend is a result of
substructures requires some assumptions about the form of the host
lens.  When these assumptions are made, the mass scale for the
substructures is very low ($\simeq 10^6\msun$) and the probable number
density is higher than expected in the $\Lambda$CDM model
\markcite{2002ApJ...580..696M}({Metcalf} 2002).  A less ambiguous system of this type
could be extremely useful for studying substructure.

\subsection{other lenses}
\label{sec:other-lenses}

In addition to the above cusp caustic cases there is 1RXS J1131-1231
which has been observed by \markcite{2003A&A...406L..43S}{Sluse} {et~al.} (2003) in V-band, but not
yet at radio wavelengths.  The cusp caustic relation is significantly
violated in this case ($R_{\rm cusp}=0.355\pm0.015$ and image opening
angle of $43.0^o$), but since microlensing by stars in the lens galaxy
could be important in the visible I choose not to emphasize this
case.  It is interesting that $R_{\rm cusp} > 0$ as expected from the
substructure hypothesis.

There are a couple of other relevant cases of spectroscopic
gravitational lensing observations.  \markcite{astro-ph/0307147}{Wisotzki} {et~al.} (2003) have shown that the
equivalent widths of the broad lines of HE0435-1223 are different in
the different images.  They attribute this to microlensing of the
optical continuum emission.   Interestingly, they still have difficulty fitting the
broad-line flux ratios to a simple lens model.  Since the
narrow-line, radio or mid-IR flux ratios are not known in this case
it is not possible to determine if larger scale substructure is
responsible for this discrepancy.  In lens SDSS J1004+4112,
\markcite{2004ApJ...610..679R}{Richards} {et~al.} (2004) have observed changes in the \civ line
profiles over a 322 day period that are not reproduced in all the
images.  They attribute this to microlensing of part of the broad line
region by ordinary stars.  Although intrinsic time variations are not yet
completely ruled out as a cause of the variations, microlensing of the
broad line region is 
particularly likely in this case because the QSO is under luminous and
thus the broad line region is relatively small.
\markcite{2004ApJ...610..679R}{Richards} {et~al.} (2004) also find time independent differences in
the \civ profiles which could be caused by some larger scale
substructure.  Further observations of a larger emission regions would
also be very revealing in this case.

It is has been shown that in general the magnification ratios of
gravitational lenses do not agree with simple lens models
\markcite{2002ApJ...567L...5M,Dalal2002}({Metcalf} \& {Zhao} 2002; {Dalal} \& {Kochanek} 2002).  \markcite{KD2003}{Kochanek} \& {Dalal} (2004) showed
that the negative magnification images tend to have smaller absolute
magnifications than are predicted by simple lens models as is expected if
substructures are causing the disagreements.  The existence of this
asymmetry is further supported by the fact that all of the observed
$R_{\rm cusp}$ quoted above are greater than zero.  The asymmetry for
$R_{\rm cusp}$ is
more extreme than it is for the distribution of just magnification ratios
between positive and negative magnification images.  Although the evidence
is pretty good that these anomalies are caused by substructures, any
constraints on the mass and density of the substructures
derived from these cases is predicated on the host lens model that is
assumed.  \markcite{EW03}{Evans} \& {Witt} (2003) showed
that some of the anomalies in non-cusp caustic cases can be explained
by adding relatively large scale axial modes to the lens models.  These models
may not be consistent with what is expected from other observations
of galaxies and their halos, but they do illustrate the ambiguities
that are inherent in deducing properties of the substructures from
simple anomalies in the magnification ratios (not to be confused with the
differential magnification ratios that are less ambiguous).

\section{Simulations}
\label{sec:simulations}

Numerical simulations are performed to calculate the image
magnification distributions.
Analytic methods for calculating these distributions are discussed
in section~\ref{sec:comp-with-cross} where it is argued that they are
not adequate for calculate the expected influence of small scale structure in the
$\Lambda$CDM model.  In this section the methods used in the
simulations are briefly described.

%% Analytic approximations can be useful when all of
%% the following conditions hold: the source is very small relative to any structure in the lens, the
%% lensing is dominated by the host and a single subhalo and the subhalo
%% is small relative to the host so that the deflection caused by the
%% host can be approximated locally by a second order expansion
%% \markcite{2003ApJ...584..664K}(see {Keeton} 2003).   
%% Generally however, there are multiple small halos affecting a single
%% image, the size of the source (in the radio, mid-IR or narrow
%% lines) is significant compared to the sizes of the substructures and
%% the effect of a single substructure on multiple images must be considered.

Any massive object near the line of sight inside or outside of the primary lens
could potentially contribute to the lensing signal.  A plane
approximation is used where the deflections caused by each object are
treated as if they take place suddenly in the plane of that lens and
the light follows an unperturbed geodesic between them.
This is known to be a very good approximation.  Given the angular
position of a point on the source, $\vec{\beta}$, the simulations must calculate the
image points, $\vec{\theta}$, that correspond to it.
If there are $N$ lenses these angular positions are related by
\begin{eqnarray}\label{lens_eq}
D_s \vec{\beta}= \vec{x}_{N+1}(\vec{\theta}) ~~~~~~~~
\vec{x}_{j+1}=D_{j+1}\vec{\theta} - \sum_{i=1}^j
D_i,_{j+1}\hat{\alpha}_i(x_i)
\end{eqnarray}
where $D_i$ is the angular size distance to the $i$th lens, $D_i,_j$ is the
distance between the $i$th and the $j$th lens planes and $D_s=D_{N+1}$ is the
distance to the source.  The deflection angle caused by the $i$th
lens is $\hat\alpha_i(\vec{x}_i)$.
Equation~(\ref{lens_eq}) is only valid for a flat cosmology because
it assumes that $D_i,_{i+1}+D_{i+1},_{i+2}=D_i,_{i+2}$.  We assume
that this is the case in this paper.

The large number of small halos and the large range in size scales,
from the size of the primary lens ($\sim 100\kpc$) to the size of the source
($\simlt 0.1\pc$ for the broad line emission region), make finding the images
and calculating their sizes challenging and time consuming.  An
adaptive mesh refinement 
technique is used to overcome these problems.  First,
equations~(\ref{lens_eq}) are solved on a coarse grid.  Minima in
$|\vec{\beta}-\vec{\beta}_s|$ are found where $\vec{\beta}_s$ is the
position of the center of the source.  The grid regions are then
modified to surround the minima.  They can be
modified in five different ways: 1) the center of the region can move,
2) it can expand or contract depending on whether the image is found
to intersect with the border of the region, 3) the grid spacing can
be made finer, 4) regions that are close together can join to become
one region, and 5) if further refinement of the grid 
fails to reach sufficient accuracy, the region can be subdivided into nine
equal subregions and the regions that do not contain any of the image
are discarded.  These modifications in the grid regions are continued
until an estimated fractional accuracy in the area of each image
reaches $10^{-4}$ or smaller.

The code is tested by comparison with several simple cases that are
solvable analytically.  The simplest is to place a substructure and
finite sized source in the center of the simulated region along with
a external shear and/or a uniform background surface density.  The
code reproduces the analytic solutions for a point mass and a untruncated Singular
Isothermal Sphere (SIS).  Further tests are discussed in
section~\ref{sec:comp-with-cross}. 
All simulations were done on a beowulf computer cluster at
the University of California, Santa Cruz. 

The entire lens is simulated at once in all cases.  However, when the
mass of the substructures is small, their number density can be very
large slowing the code down.  To reduce this problem the angular range for
the positions of substructures included in the calculation is made
smaller for smaller masses.  Limiting the range appears to have a
small effect  on the results ($\simlt 10\%$ on the magnification) if
the region is kept large enough to contain over 150 subhalos per decade in mass.

For intergalactic halos the Press-Schechter formalism
\markcite{1974ApJ...187..425P}({Press} \& {Schechter} 1974, with the extra factor of 2) is used to
calculate the mass function from which a random sample of halos is
drawn.  It is known that the modified mass function of
\markcite{2002MNRAS.329...61S}{Sheth} \& {Tormen} (2002) is a better fit to cosmological Nbody
simulations.  However, these mass functions are very close to
each other for the range of halo masses used in this paper.  Although
the Press-Schechter mass function underestimates the number a halos in
the mass range $10^{10}-10^{13}\msun$, for $m\simlt
10^{10}\msun$ the Sheth-Tormen mass function is larger.  The two mass
functions are within a factor of 1.3 for $10^6\msun \simlt m \simlt
10^{10}\msun$.  Using the Press-Schechter mass function is more
conservative. 
The structure of these halos is
taken to be of the NFW form truncated at the virial radius.  The
initial power spectrum is taken to be scale invariant and normalized
to $\sigma_8=0.9$.  The concentrations of the halos are set according
to equation~(\ref{concentration}).

The subclumps inside the primary lens are treated as a different
population and calculated in separate simulation runs since their
abundance is considerably less certain.  They are also
of the NFW form, but they are truncated.  The
truncation is done by using the standard approximation to the tidal radius 
\begin{equation}
r_t(m,R) = \frac{R~ m^{1/3}}{\left( M(R) [3-\alpha] \right)^{1/3}}~;~\alpha\equiv
\frac{\partial\ln M}{\partial\ln R}
\end{equation}
where $R$ is the distance from the center of the host halo, $M(R)$ is
the mass of the host within that distance and $m$ is the mass of the subhalo.
For the purposes of calculating the tidal radius, the host is taken to
be a SIS and $R$ is set to 5 times the Einstein radius of the host lens.  Since not much is known about the
mass function of substructures inside a host halo in this mass range, it
was decided to use substructures of just one mass at a time and adjust the total
surface density of them.  This makes interpretation of the results
more straightforward.

In addition to the substructure, a model for the host lens must be
chosen.  The substructure will change the positions of the images
slightly so if a lens model is chosen to fit the observed image
positions perfectly it will not fit them perfectly after the
substructure is added.  To produce a
perfectly consistent lens model one would have to adjust the host lens
model for each realization of the substructure.  This is
very computationally expensive and not necessary in practice.
The shifts in positions are generally small when the masses of the
substructures are small ($\simlt 0.1\arcsec$ for $m<10^8\msun$) and, in addition, since the host lens
model is degenerate it is ambiguous how it should be adjusted to
correct for the shift.  The goal here is to reproduce all the significant
characteristics of the observed lens -- image configuration, rough
image opening angle (within $2^\circ$), redshifts of source and lens -- so that one can
determine whether lenses the look like the ones observed and have the
observed ratio anomalies are common in the $\Lambda$CDM model.  

Often, when the host lens model is set up to produce an extreme cusp
or fold caustic configuration and the substructure includes masses of
$\simgt 10^8 \msun$, the image configuration will be changed so that
two of the images are no longer present.  In the statistical studies
presented in section~\ref{sec:results} these cases are simply ignored
on the basis of their being incompatible with the lens systems that
are being modeled.

\subsection{comparison with cross section method}
\label{sec:comp-with-cross}

Another way to calculate the magnification probability distribution
is to use a cross section or optical depth method.  This approach
significantly reduces the computational work necessary.  
The magnification as a function of source position is 
calculated for one subhalo and the host lens is represented by
 a constant background shear and smooth surface density.  From this
 the cross section, $\sigma(\mu)$, is found for a single halo. 
The probability of getting a magnification of above $\mu$ is then
approximated as $\tau(\mu) = \eta\,\sigma(\mu)$ where $\eta$ is the
number density of halos.  \markcite{2003ApJ...584..664K}{Keeton} (2003) found an analytic
approximation to $\sigma(\mu)$ for a untruncated SIS halo and a
point-like source.  \markcite{2003ApJ...592...24C}{Chen}, {Kravtsov}, \&  {Keeton} (2003) used this result to
estimate the influence of halos inside and outside of the primary lens
on QSO magnification anomalies.

This method has several drawbacks.  Firstly, the cross section
approach is only valid for rare events.  A typical source will lie
roughly equidistant between subhalos; the assumption that the
magnification can be calculated using a single subhalo will be valid
in only a minority of cases.  This would be acceptable if the
subhalos where very spars and the majority of sources were unaffected
by them.  This is generally not the case however.  A typical line of
sight out to $z=2$ passes within one virial radius of approximately 300
halos with masses between $10^7\msun$ and $10^9\msun$.  The standard
deviation in the total convergence and shear caused by these halos is on the
order of several percent which will cause changes in the
magnifications of the sizes seen in the these simulations \markcite{M04b}(see Metcalf 2004).

Another problem is that to simplify the
calculation in this method the source is taken to be infinitely small compared to
the subhalo.  This is not a good approximation for a radio or mid-IR
source which can be larger than the subhalos under consideration.
The subhalo is also taken to be untruncated which will clearly not be the
case, especially near the center of the primary lens.  Even if the
lensing were not sensitive to the truncation radius, it would affect the
conversion between subhalo number density and mass density.

The cross section method can also breakdown because the deflections
caused by the primary lens are not well approximated by a simple shear and
convergence.  This is particularly true near the caustics which are of
special interest here.  Also, the substructure is taken 
to influence each image independently in the cross section approach.
This is not always the case, especially when the images are close
together.  For example, a single subhalo, if large enough ($\sim 10^{10}\msun$) can shift
the position and shape of a cusp caustic, changing the positions of
all three close images.

The analytic cross section of \markcite{2003ApJ...584..664K}{Keeton} (2003) has been
used to test the lensing code used in this paper.  The code should
return the same results when there is only one untruncated subhalo and
the source is made very small.  This was done for the local shear and
convergences appropriate for the 4 images of Q2237+0305.  Good agreement was
found for $\Delta\mu > 0.03\mag$ below which the finite size of
simulated region became important.
     
\section{Results}
\label{sec:results}

\begin{figure}[t]
\centering\epsfig{figure=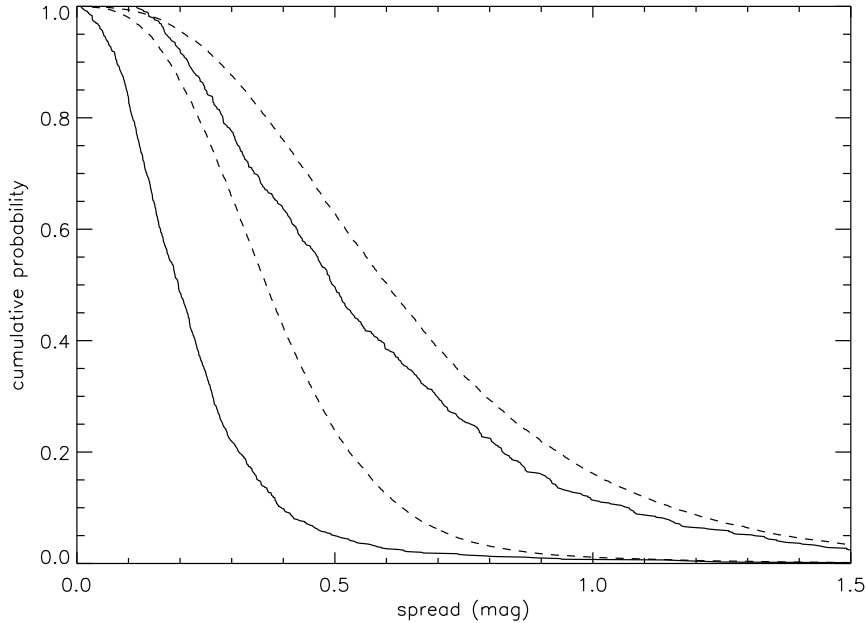,height=3.5in}
\caption[the]{\footnotesize This is the probability of having
a magnification ratio disagree with the lens model by more than a
certain magnitude for Q2237+0305.  The two solid curves are without
observational noise and the dashed curves are with 0.15~mag of noise.
For each type of curve the one on the left is for intergalactic halos
with $10^7\msun < m < 10^8\msun$ and the one on the right is for
$10^7\msun < m < 10^9\msun$.  There is no substructure inside the
primary lens.}
\label{fig:spread_ratio}
\end{figure}

\begin{figure}[t]
\centering\epsfig{figure=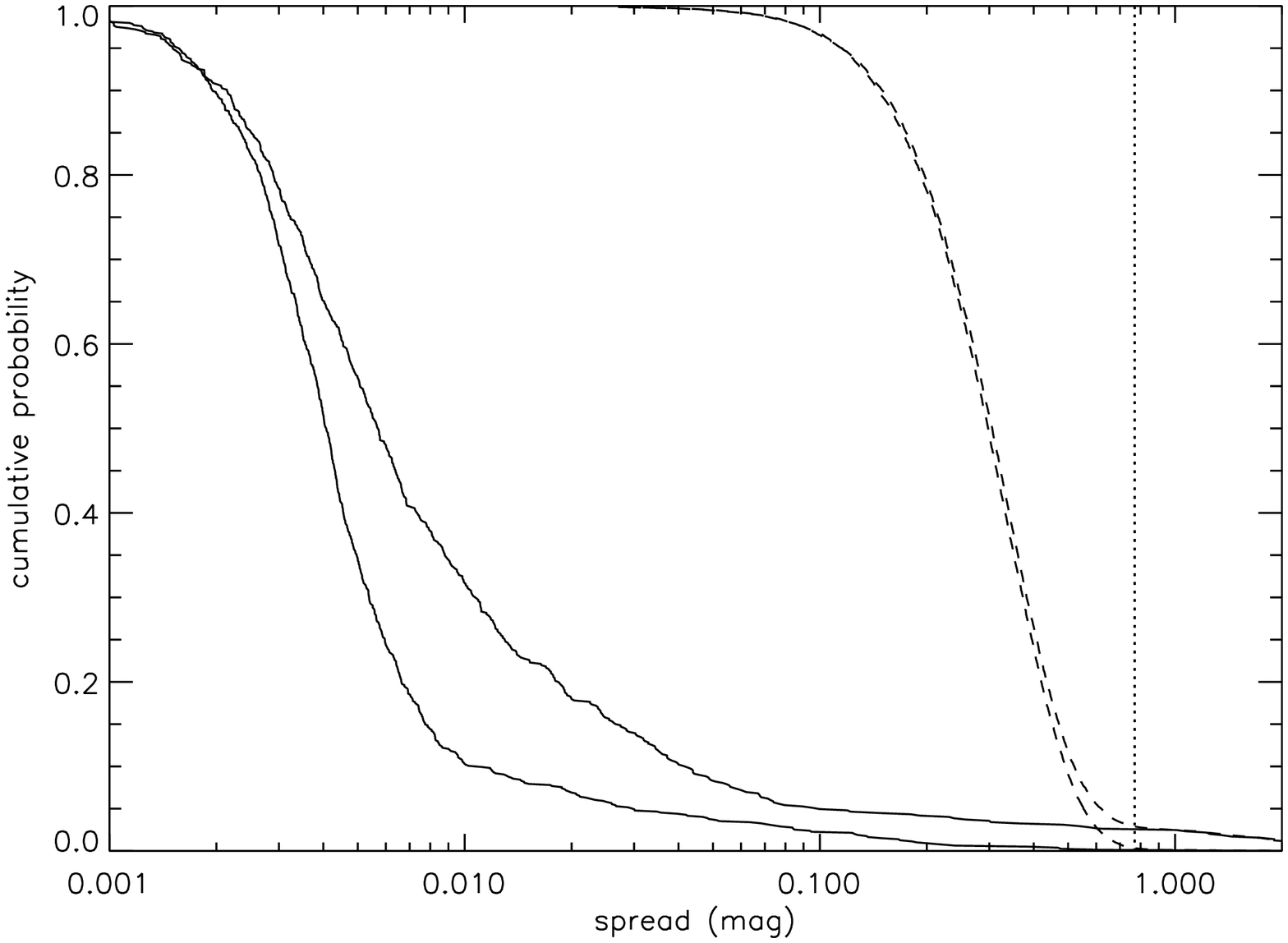,height=3.5in}
\caption[the]{\footnotesize The solid curves are the probability of the
  spread in the differential magnification ratios being above a certain
  magnitude for Q2237+0305.  The source sizes used are 1\pc and
  100\pc.  The substructure mass ranges are $10^7\msun < m < 10^8\msun$ for the
  left most curve and $10^7\msun < m < 10^9\msun$ for the solid curve
  on the right.  The dashed curves are the same but with 0.15~mag 
  of noise which was the level measured in \markcite{cirpass2237}{Metcalf} {et~al.} (2004).  The
  dotted line marks the measured discrepancy between the
  radio/mid-IR magnification ratios and the narrow line magnification
  ratios reported in that paper. }
\label{fig:spread_dmr}
\end{figure}

Simulations were performed to mimic the observed lenses discussed in
section~\ref{sec:summary-observations} with the addition of
$\Lambda$CDM substructure.  The resulting combinations of image
magnifications are then compared with those observed to determine if
the observed anomalies are expected to be reasonably common in this
cosmological model.

To represent lens Q2237+0305, and other lenses in the Einstein cross
configuration, a host lens model is constructed that fits
the image positions of Q2237+0305.  The model consists of a Singular
Isothermal Ellipsoid (SIE) with an external shear and fits the image
positions very well.  The effects of substructure within the host lens
and its contributions to spectroscopic lensing were investigated in
\markcite{cirpass2237}{Metcalf} {et~al.} (2004).  Just the intergalactic contribution is discussed
here.  All the halos within 2~arcsec of the center of the lens are
included in the simulations.

For each realization of the substructure the three magnification
ratios can be compared with the ratios expected from the host lens model.
Figure~\ref{fig:spread_ratio} shows a cumulative distribution of the
largest discrepancy (in magnitudes) out of these three between the
model and simulated values.  The source size is 1~pc in this case.  
By comparing the curves figure~\ref{fig:spread_ratio} it is found that
most of the anomalies are caused by the high end of the mass 
distribution, $m\simeq 10^8-10^9\msun$.  
One can see that these discrepancies are
rather large even without any substructure in the host lens itself.
Discrepancies as large as $\sim 0.5\mag$ are expected in half the
cases.  The typical discrepancies between observed flux ratios and
models are a few tenths of a magnitude
\markcite{2002ApJ...567L...5M,KD2003}(see {Metcalf} \& {Zhao} 2002; {Kochanek} \& {Dalal} 2004).  This makes the observed
ratio anomalies consistent with $\Lambda$CDM, simple lens models and
no substructure internal to the primary lenses.  

Although figure~\ref{fig:spread_ratio} demonstrates a consistency with
$\Lambda$CDM it is not certain that CDM substructures are the only
possible explanation for the discrepancies 
in Einstein cross lenses.  Some of the discrepancy could be accounted
for by a less than perfectly symmetric host lens.  Although this
probably cannot account for all of the discrepancies, it can
significantly change the amount of substructure that is required to
produce them and thus it is not a strong constraint on the $\Lambda$CDM model.

As described in section~\ref{sec:spectr-grav-lens}, a more restrictive
test comes from the spectroscopic lensing observations of Q2237+0305.
Figure~\ref{fig:spread_dmr} shows the cumulative distribution of the
spread in the differential
magnification ratios between a 100~pc source and a 1~pc source.  These
are very small; much smaller than the spread of $0.77\pm 0.15\mag$ between the
narrow line emission region and the mid-IR emission region measured by
\markcite{cirpass2237}{Metcalf} {et~al.} (2004).  CDM halos seem easily capable of changing the
magnification ratios by this much, but they do not produce the
mismatch in the magnifications of different size sources.  This
problem can be traced to a deficiency of small mass ($\sim 10^6$)
halos in the $\Lambda$CDM model.  As we shall see this is the only strong
inconsistency between the $\Lambda$CDM model and magnification ratio measurements.

\begin{figure}[t]
\centering\epsfig{figure=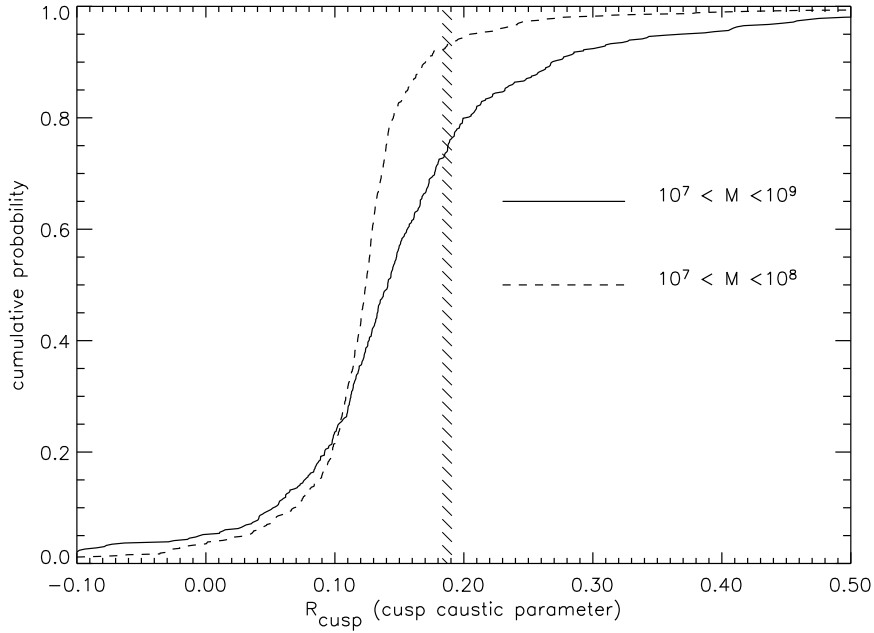,height=3.5in}
\caption[the]{\footnotesize The distribution of the cusp caustic
  parameter, $R_{\rm cusp}$, for lens B1422+231 with only
  intergalactic standard $\Lambda$CDM small-scale structure.  The
  observed value in the radio with error is shown as the hashed
  region.  The different curves correspond to the halos mass ranges
  shown.  It can be seen that most of the changes in $R_{\rm cusp}$
  are caused by relatively large mass halos, $10^8\msun<m<10^9\msun$.
  There is about a 25\% chance of $R_{\rm cusp}$ differing from zero
  by more than is observed.}
\label{fig:r_prob1422}
\end{figure}

In considering the case of B1422+231 the same kind of simulations are
performed only the cusp caustic parameter, $R_{\rm cusp}$, is calculated for each
realization.  The host lens is again a SIE+shear model fit to the
observed image positions.  Figure~\ref{fig:r_prob1422} shows the
distribution of $R_{\rm cusp}$ with the expected population of
intergalactic halos only.  The first thing to note is the marked
asymmetry in the distribution.  As previously seen
\markcite{astro-ph/0109347,2001ApJ...563....9M,2002ApJ...580..685S}(Metcalf 2001; {Metcalf} \& {Madau} 2001; {Schechter} \& {Wambsganss} 2002), the
magnifications of negative magnification images are affected by
substructure differently than positive magnification images.
When substructure is added, $R_{\rm cusp}$ should be biased toward positive
values as seen here. 

Also shown in figure~\ref{fig:r_prob1422} is the observed value of
$R_{\rm cusp}$ for comparison.  There is a perfectly reasonable
probability of $\simeq 0.28$ that $R_{\rm cusp}$ would be even
larger than the observed value.  By comparing the two different ranges
for the halo masses, it can be seen that violations in the cusp caustic
relation are mostly caused by more massive halos in this case.  Also note that a
negative $R_{\rm cusp}$ of the same magnitude would be clearly
inconsistent with this explanation.  In light of this, the violation of
the cusp caustic relation in B1422+231 seems fully consistent with
the $\Lambda$CDM model even without substructure within the halo of
the primary lens.

We can also compare figure~\ref{fig:r_prob1422} to lens B0712+472
which has a similar configuration to B1422+231 although a lower source
redshift.  It is easily seen that its value of $R_{\rm cusp}=0.26\pm
0.02$ is not particularly unlikely (there is a $\sim 12 \%$ probability of it
being larger) and thus does not require an additional
explanation beyond the expected population of intergalactic halos.
Considering the additional substructure within the host lens, the
observed $R_{\rm cusp}$ seems perfectly consistent with $\Lambda$CDM.
Although a precise calculation would require modeling this particular
lens specifically, the results would not change greatly if that was done.

\begin{figure}[t]
\centering\epsfig{figure=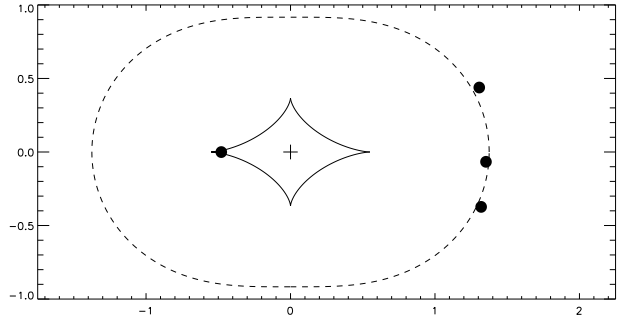,height=2.5in}
\caption[the]{\footnotesize Diagram of the close cusp caustic model used
  in simulations. The dots are where the centers of the images are and
the + marks the center of the lens.  The units are in arcseconds.  The
image opening angle is $25.5^o$.  The redshifts used are the same as
for B2045+265, $z_{\rm lens}=0.87$ and $z_{\rm source}=1.28$.}
\label{fig:diagram2045}
\end{figure}

Lens B2045+265 is a more extreme cusp caustic case.  When the source
is very near the cusp, substructure can have a significant effect on
the details of the lens configuration such as the precise image opening
angle.  After substructures are added to a host lens model, the image
positions will not fit the 
observed ones precisely, but the lens will still be very similar in
its general aspect.  To investigate the violations of the cusp caustic
relation in cases like this, a SIS+shear host lens model is constructed
that reproduces the approximate size and image opening angle of
B2045+265.  The image configuration for this model is shown in
figure~\ref{fig:diagram2045}.

\begin{figure}[t]
\centering\epsfig{figure=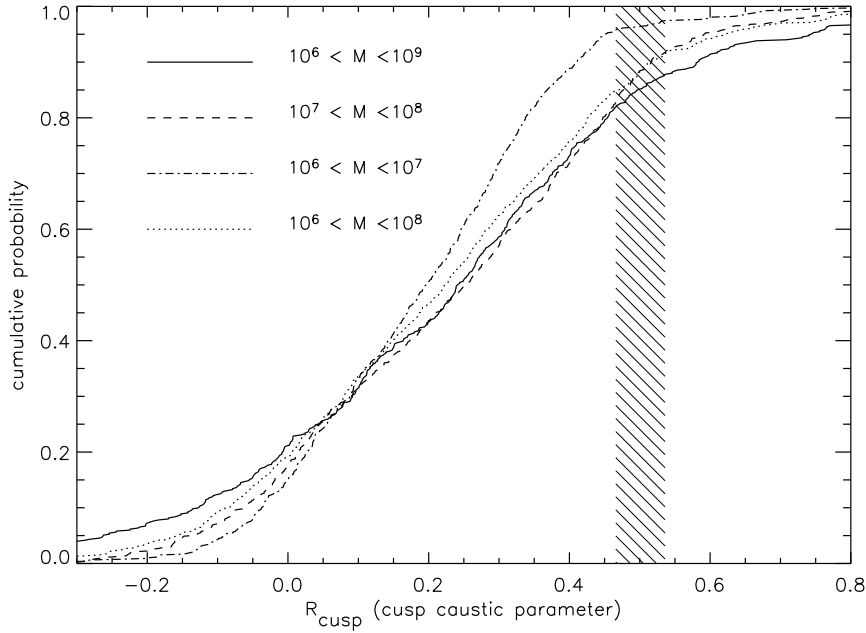,height=3.5in}
\caption[the]{\footnotesize The cumulative distribution for $R_{\rm
    cusp}$ in the tight long axis case like B2045+265 with only
  intergalactic substructure.  The observed value of $R_{\rm cusp}$ in
the radio is shown by the hashed region.  The included subhalo mass
ranges are shown.}
\label{fig:r_prob2045cosmic}
\end{figure}

Figure~\ref{fig:r_prob2045cosmic} shows the results for simulations
with just intergalactic $\Lambda$CDM halos.  Also shown is the
observed value for $R_{\rm cusp}$.  With a halo mass range of
$10^6\msun<m<10^9\msun$ the observed $R_{\rm cusp}$ does not appear
strongly disfavored -- $15\%$ chance of it being larger.
Again one sees the strong asymmetry of the distribution.  An observed
value of $R_{\rm cusp} \simlt -0.3$ would have been strong evidence
against the substructure explanation for the magnification ratio anomalies.

\begin{figure}[t]
\centering\epsfig{figure=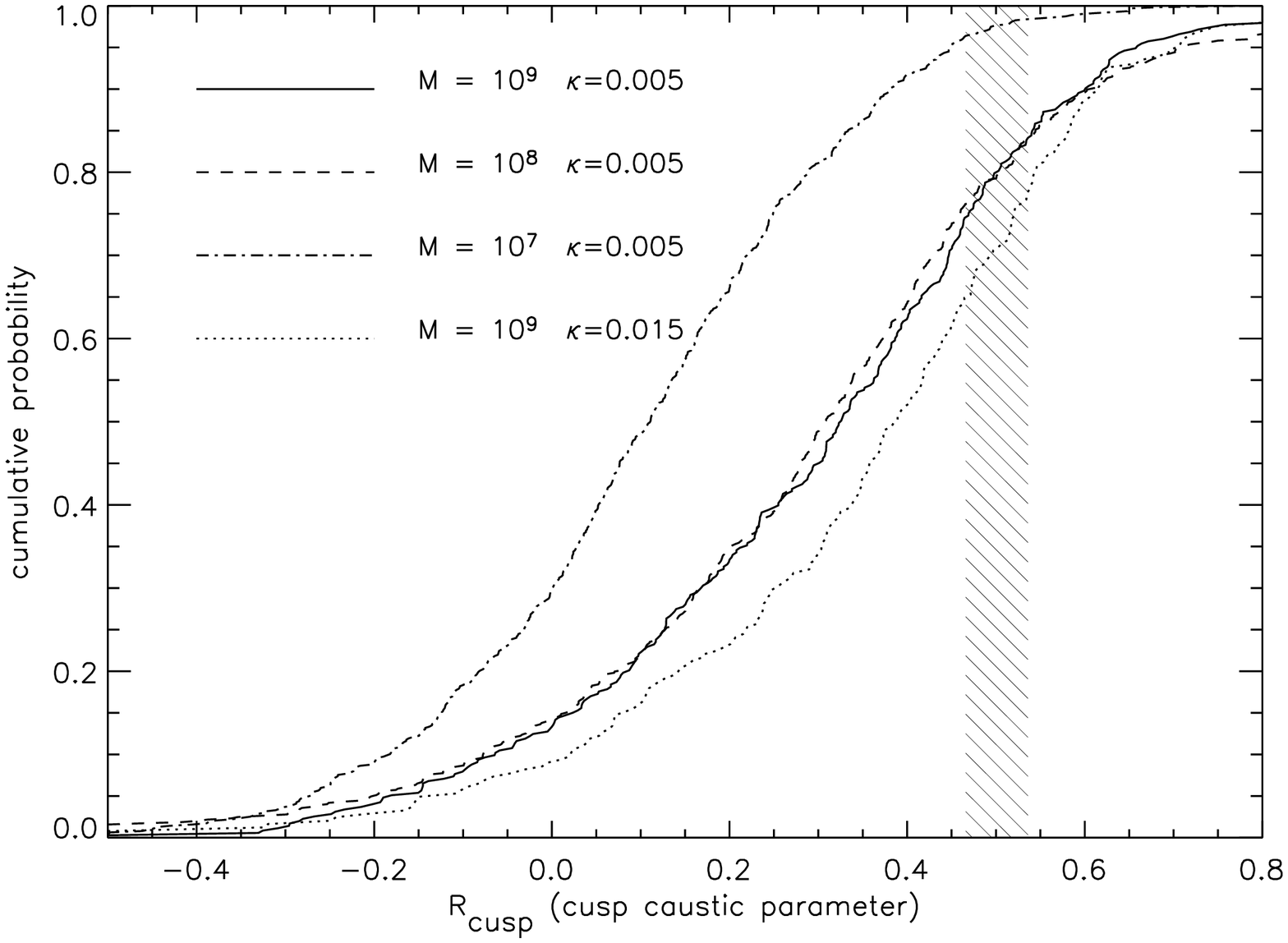,height=3.5in}
\caption[the]{\footnotesize The cumulative distribution for $R_{\rm
    cusp}$ in a tight long axis case like B2045+265 with substructure
inside the host lens.  In each case the subhalos are of the mass
as indicated.  The total average surface density in substructure is
indicated in units of the critical density.  A surface density of
$\kappa=0.005$ is approximately 1\% of the surface density.  The
observed value and errors are indicated by the hash marks.}
\label{fig:r_prob2045internal}
\end{figure}

The importance of substructures within the host lens for a
B2045+265--like lens was also investigated.  For the $10^9\msun$ and
$10^8\msun$ cases the range was 2~arcsec from the 
center of the lens.  Because of the high number of individual subhalos
in the $10^7\msun$ case the range was reduced to the 1.61~arcsecs
surrounding the image triplet.  Figure~\ref{fig:r_prob2045internal}
shows the results for different substructure masses and surface
densities.  For a host lens with a radial profile similar to a SIS
($\rho(r)\propto r^{-2}$),
the Einstein radius -- and thus the images -- forms where $\kappa\simeq
0.5$.  For this reason we expect a substructure surface density
of $\kappa=0.005$ to be $\sim 1\%$ of the total surface density in the
lens.  From figure~\ref{fig:r_prob2045internal} it can be seen that
this is enough substructure to account for the observed $R_{\rm
  cusp}$ if the mass scale is $\sim 10^8\msun$ or greater.
Substructure within the primary lens could be the most significant
cause of the anomaly in this case, but comparing
figure~\ref{fig:r_prob2045cosmic} to
figure~\ref{fig:r_prob2045internal} shows that the contributions from
internal and  external substructure are comparable.

The importance of intergalactic halos will come as a surprise to
some.  Calculating some simple numbers can make it less so.  The
total $\kappa$ (surface density weighted by the critical density)  in
halos below $10^9\msun$ along a line of sight to 
$z=2$ is $\sim 0.15-0.19$.  The variance in this number is
$\langle\kappa^2\rangle^{1/2}\simeq 0.04$ with the halo model used
here.  This is close to $10\%$ of the surface density of the primary
lens, larger than the expected level needed to cause the monochromatic
magnification anomalies.

\section{Discussion}
\label{sec:discussion}

It has been shown here that anomalies in the monochromatic (as opposed
to differential) magnification ratios of
cusp caustic lenses might be explained naturally within the
$\Lambda$CDM model with little if any substructure within the dark
matter halo of the primary lenses.  Intergalactic halos could be enough to
account for these anomalies.  This conclusion is derived from
simulating several realistic and representative cases where it is
shown that the cusp caustic relation is violated by such halos.
Furthermore, the typical observed 
anomalies in the monochromatic magnification ratios of 
several tenths of magnitudes  -- when compared to simple lens models
-- are easily explained in the same way.  
The contribution to flux anomalies from intergalactic halos is
found to be significant.  Measuring the amount of substructure that is within
the primary lens halos for comparison with Nbody simulations will
require a large number of lenses and an 
accurate prediction for the intergalactic contribution.  
These anomalies in the monochromatic magnification ratios could also
be explained by smaller scale structures since they do not provide
significant constraints on the substructure mass.  The fact that all of
the observed cusp caustic parameters, $R_{\rm cusp}$, are positive is
further support for the conclusion that these anomalies are being caused by
some kind of substructure.

The alternative to intergalactic halos, substructure in the primary
lens, could also be contributing to the magnification ratio anomalies
although the expected abundance of such substructures is not yet
certain.  \markcite{2004ApJ...604L...5M}{Mao} {et~al.} (2004) have argued that Nbody
simulations indicate that there is not enough substructure in
$\Lambda$CDM halos to explain the lensing observations.  This argument
requires extrapolating the mass function of subhalos beyond the
limitations of the current simulations to smaller masses
and further into the centers of the halos.  For this reason, it
cannot yet be determined if the additional intergalactic halos cause
magnification anomalies to be overabundant relative to observations.

\markcite{2003ApJ...592...24C}{Chen} {et~al.} (2003) found that intergalactic halos play a
significant, but less important role in the magnification anomalies.
The disagreement with this paper appears to be a result of
\markcite{2003ApJ...592...24C}{Chen} {et~al.} (2003) not taking into account of
deflections by multiple halos and approximating the host lens as a simple
shear and constant surface density instead of modeling it in more
detail (see section~\ref{sec:comp-with-cross}).   The collective
surface density in small, intergalactic halos is significant and
varies across the sky.  These perturbations in the surface density are
enough to change the image magnifications by tenths of a percent.

In contrast to the monochromatic magnification ratios, the
spectroscopic gravitational lensing observations of Q2237+0305
require more small mass halos than are expected in the $\Lambda$CDM model.
Bent multiply imaged radio jets also hint, although less securely, at
a large number of small mass objects \markcite{2002ApJ...580..696M}({Metcalf} 2002).  The
case for small mass substructure is not yet secure, but further data should resolve the
issue.  On the theoretical side, advances in cosmological simulations
should soon make it possible to extend predictions for the mass function
of substructures within the halos of large galaxies down to smaller
masses and smaller galactocentric radii where they can be more
directly compared with observations.   At this time, there is an
inconsistency between the $\Lambda$CDM model and the gravitational
lensing observations that needs to be resolved.

\acknowledgments
The author would like to thank J. Bullock for very useful discussions and
M. Magliocchetti for very helpful suggestions.  I would also like to thank
J. Primack and his group for allowing me to use their beowulf
computer cluster.
Financial support was provided by NASA through Hubble Fellowship 
grant HF-01154.01-A awarded by the Space Telescope Science Institute,
which is operated by the Association of Universities for Research 
in Astronomy, Inc., for NASA, under contract NAS 5-26555

%\bibliographystyle{/Users/bmetcalf/TeX/astronat-1.6/apj/apj}
%% \bibliography

%%\bibliographystyle{/u/bmetcalf/TeX/astronat-1.6/apj/apj}
%\bibliography{}

\end{document}